\documentclass{emulateapj}
\usepackage{apjfonts}

\slugcomment{Accepted by the Astrophysical Journal Letters}

\shorttitle{IR spectra of Oph\,162225-240515}
\shortauthors{Brandeker et al.}

\begin{document}

\title{Infrared Spectroscopy of the Ultra Low Mass Binary Oph~162225-240515}

\author{Alexis Brandeker, Ray Jayawardhana}
\affil{Department of Astronomy and Astrophysics, University of Toronto,
    50 St.~George Street, Toronto, ON M5S~3H4, Canada}


\author{Valentin D.\ Ivanov}
\affil{European Southern Observatory, Ave.\ Alonso de Cordova 3107, Vitacura, 
Santiago 19001, Chile}

\and 

\author{Radostin Kurtev}
\affil{Departamento de F\'{i}sica y Meteorolog\'{i}a, Facultad de Ciencias,
Universidad de Valpara\'{i}so, Av.\ Gran Breta\~{n}a 644, Playa Ancha, Casilla 53, Valpara\'{i}so, Chile}

\begin{abstract}
Binary properties are an important diagnostic of the star and brown dwarf formation processes. While wide binaries appear to be rare in the sub-stellar regime, recent observations have revealed Ophiuchus~162225-240515 (2MASS J16222521-2405139) as a likely young ultra-low-mass binary with an apparent separation of $\approx$240 AU. Here, we present low-resolution near-infrared spectra of the pair from NTT/SOFI ($R\sim600$) and VLT/ISAAC ($R\sim1400$), covering the 1.0--2.5\,$\mu$m spectral region. By comparing to model atmospheres from Chabrier \& Baraffe and Burrows et al., we confirm the surface temperatures to be $T_{\mathrm{A}}=(2350\pm150)$~K and $T_{\mathrm{B}}=(2100\pm100)$~K for the two components of the binary, consistent with earlier estimates from optical spectra. Using gravity sensitive K\,I features, we find the surface gravity to be significantly lower than field dwarfs of the same spectral type, providing the best evidence so far that these objects are indeed young. However, we find that models are not sufficiently reliable to infer accurate ages/masses from surface gravity. Instead, we derive masses of $M_{\mathrm{A}} = 13^{+8}_{-4}$\,M$_{\mathrm{J}}$ and $M_{\mathrm{B}} = 10^{+5}_{-4}$\,M$_{\mathrm{J}}$ for the two objects using the well-constrained temperatures and assuming an age of 1--10\,Myr, consistent with the full range of ages reported for the Oph region.
\end{abstract}

\keywords{binaries: general --- stars: individual (2MASS J16222521-2405139) --- stars: low-mass, brown dwarfs --- stars: pre-main sequence --- stars: planetary systems}

\section{Introduction}
Growing evidence suggests that the bottom end of the stellar initial mass function extends well into
the giant planet regime ($<$15\,M$_{\mathrm{J}}$). A few dozen free-floating brown dwarfs with inferred masses
near or below the deuterium-burning limit have been identified in the $\sigma$~Orionis association
\citep{zap00} and the Orion Nebula Cluster \citep[ONC,][]{luc00}. Some of them have
been confirmed as cool objects that are likely young cluster members through follow-up spectroscopy
\citep{mar01,bar01,luc01,luc06}.

However, very little is known about the properties of these isolated planetary mass objects (IPMOs or 
`planemos') because they are extremely faint at the distances of $\sigma$\,Ori and ONC ($\sim$350--450 pc),
and only a couple more have been identified in closer star forming regions \citep{tes02,luh05}. Recently, 
by combining ground-based optical and near-infrared photometry with
Spitzer Legacy Survey data, \citet{all06b} identified 6 new candidate planemos in three nearby
regions, at distances $\lesssim$150 pc. What's more, based on their infrared excess, these objects appear
to be surrounded by circum-sub-stellar disks, just like many of the higher mass young brown dwarfs
\citep[e.g.,][]{jay03}. Based on optical spectra, \citet{jay06b} confirmed four
of the candidates as ultra low mass objects and a fifth as a somewhat higher mass brown dwarf; the sixth
turned out to be a likely background source.

\citet{jay06c} reported that one of the newly identified `planemos', Oph\,162225-240515 (hereafter Oph\,1622),
is in fact a $\sim$240 AU binary, based on optical and infrared images as well as optical
spectra.
Independently, \citet{all06a} and later \citet{clo06} also observed 
Oph\,1622 to be binary.
By comparison to late-type objects, \citet{jay06c} derived spectral types of M9 and
M9.5--L0, for the primary and the secondary, respectively. The authors made a strong case for the youth,
coevality and physical association of the two objects, based on several lines of evidence, and derived
masses of $\sim$14\,M$_{\mathrm{J}}$ and 7--8\,M$_{\mathrm{J}}$ for them in comparison to \citet{bar03} models.

The existence of an ultra low mass binary with a wide separation comes as somewhat of a surprise, given
the paucity of wide binaries in the sub-stellar regime \citep[e.g.,][]{bou03,giz03,krau06}, and poses 
a challenge to the ejection model for brown dwarf
formation \citep[e.g.,][]{bat02b}. Given the importance of Oph\,1622 as a benchmark binary, here
we present follow up infrared spectroscopy to derive independent additional constraints on the nature of 
this pair.

\section{Observations}

Spatially resolved, low-resolution ($R$$\sim$1400) near-infrared (NIR) spectra of Oph\,1622 
(2MASS J16222521-2405139) were obtained 
in service mode under excellent seeing ($<$0\farcs6 in $V$) on 2006 July 11 in the $J$- and $K$-band, and
on 2006 July 19 in the $H$-band. The NIR instrument ISAAC was used in spectroscopic mode, mounted on the
8.2\,m Antu telescope, one 
of the Very Large Telescope units located at Paranal, Chile, operated by the European Southern Observatory. 
A slit width of 0\farcs3 was used, and a pixel scale of 0\farcs147\,pix$^{-1}$ on a 1024$^2$ pixel array. 
The spectral regions covered are 1.07--1.44\,$\mu$m, 1.42--1.89\,$\mu$m, and 1.84--2.56\,$\mu$m, for $J$,
$H$, and $K$, respectively. The slit was aligned with the binary and the telescope nodded 
15\arcsec\ along the slit direction between exposures. Exposure times were 8$\times$120\,s in $J$,
 20$\times$120\,s in $H$, and 30$\times$117\,s in $K$. As telluric standards, the B stars HIP\,84292 and 
HIP\,84435 were observed on 2006 July 11 and 2006 July 19, respectively.

Additional lower resolution spectra ($R$$\sim$600), covering 0.95--1.64\,$\mu$m and 1.53--2.50\,$\mu$m, 
were obtained under average seeing ($\sim$1\farcs1 in $V$) on 2006 August 10 with the SOFI instrument on 
the 3.58\,m New Technology Telescope, located at La Silla, Chile, and operated by the European Southern 
Observatory. A 1\arcsec\ slit was used, and a pixel scale of 0\farcs288\,pix$^{-1}$ on a 1024$^2$
pixel array. The slit was aligned with the binary and the telescope nodded 
25\arcsec\ along the slit direction between exposures. Exposure times were 2$\times$120\,s for both the 
blue and red grisms. The sun-like G star HIP\,84181 was observed as a telluric standard.

Reduction of the ISAAC and SOFI spectra proceeded in a similar fashion. The sky for every exposure was 
estimated by averaging over temporally close exposures, were the star was offset to a different position 
on the array. After sky subtraction, the frames were flatfielded. The spectra were then extracted using
an optimal extraction algorithm developed by Marten H.\ van Kerkwijk (private comm.), that fits two
Moffat functions \citep{mof69} simultaneously to the binary. The $\sim$0\farcs5 and 1\farcs1 resolutions of
ISAAC and SOFI, and the low flux ratio ($\sim$2) and large separation ($\sim$2\arcsec) between the components, 
ensure that contamination between
the two extracted spectra is not an issue. Wavelength calibration was obtained by measuring telluric
OH emission lines \citep{rou00} found in abundance in the science frames. The procedure was repeated for 
the telluric standards. The science spectra were then flux calibrated by multiplying by a model spectrum
of the standard star \citep[from][]{pic98}, resampled to the observed sampling, and dividing by the 
observed standard. No extinction correction was made to the spectra.

\section{Results and discussion}

\begin{figure}
\includegraphics[width=\hsize,clip=true]{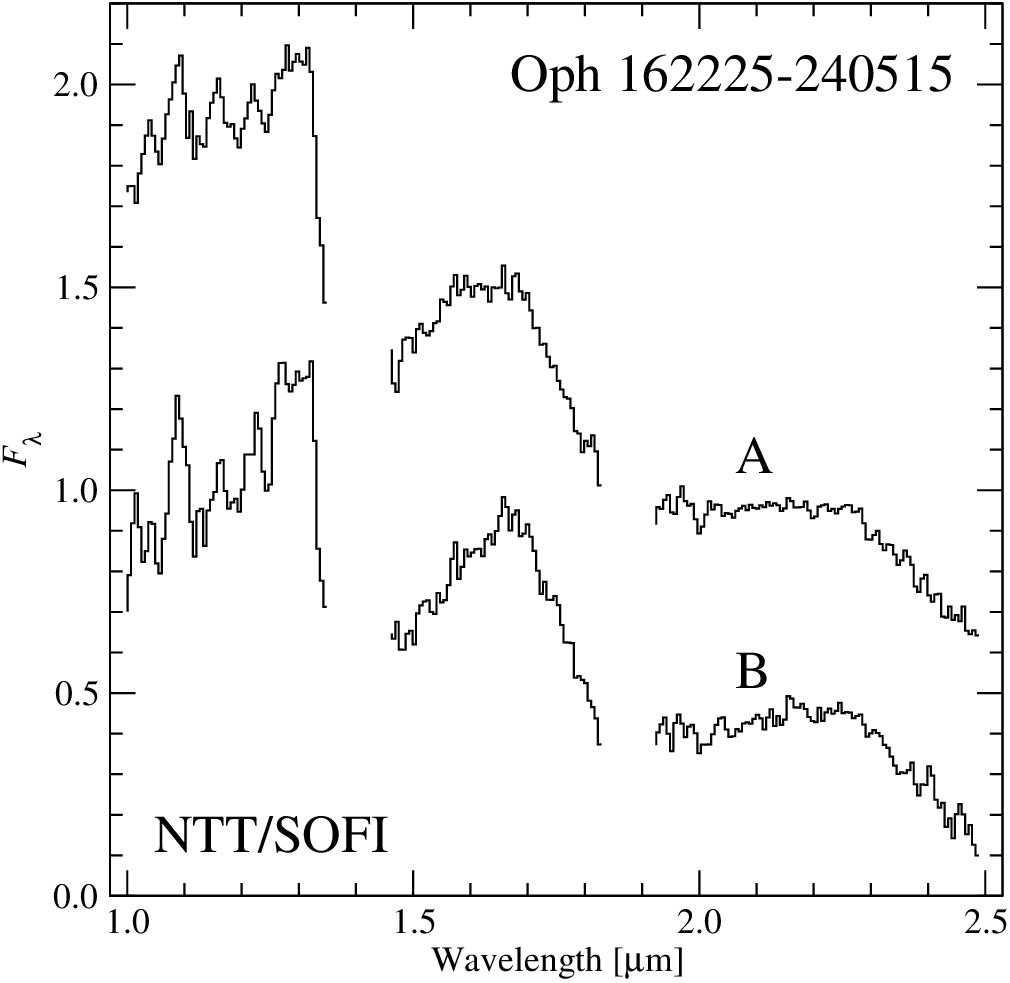}
\caption{\label{f:sofi}SOFI spectra of Oph\,1622 A and B, rebinned to a resolution of 
 $R$$\sim$150 (60\,\AA\,pix$^{-1}$). The spectra have been normalized to 1 at 1.65\,$\mu$m,
and the A spectrum offset by 0.5.}
\end{figure}

\begin{figure}
\includegraphics[width=\hsize,clip=true]{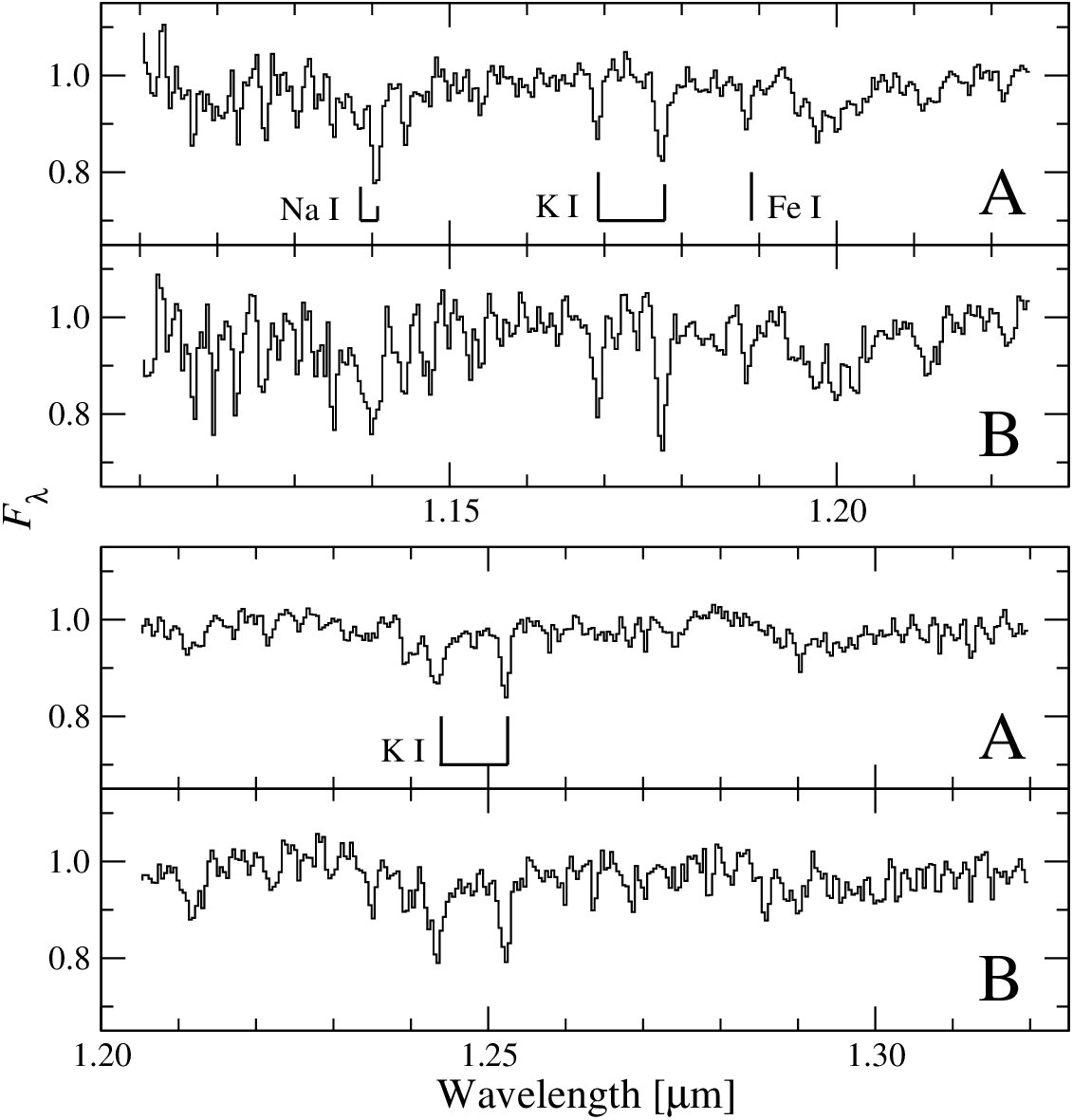}
\caption{\label{f:J} ISAAC $J$-band spectra of Oph\,1622 A and B, divided by the (pseudo-)continua. 
Several atomic features are identified. 
 }
\end{figure}

\begin{figure}
\includegraphics[width=\hsize,clip=true]{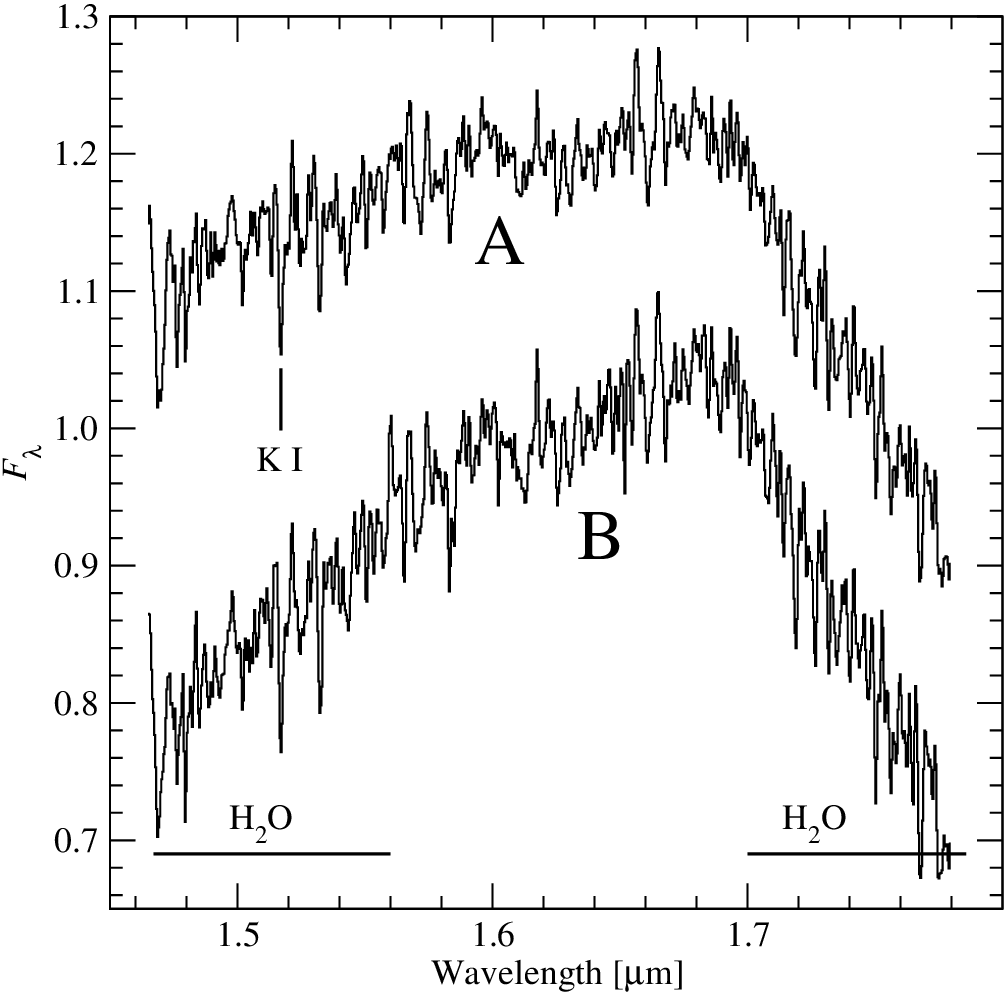}
\caption{\label{f:H} ISAAC $H$-band spectra of Oph\,1622 A and B. The spectra have been normalized
to 1 at 1.65\,$\mu$m, and the spectrum of A offset by 0.2. In addition to the
water vapour suppression characteristic for late-type, low-gravity objects, a
K\,I absorption is visible at 1.52\,$\mu$m. Many of the other features can be
attributed to FeH absorption \citep{cus03}.}
\end{figure}

\begin{figure}
\includegraphics[width=\hsize,clip=true]{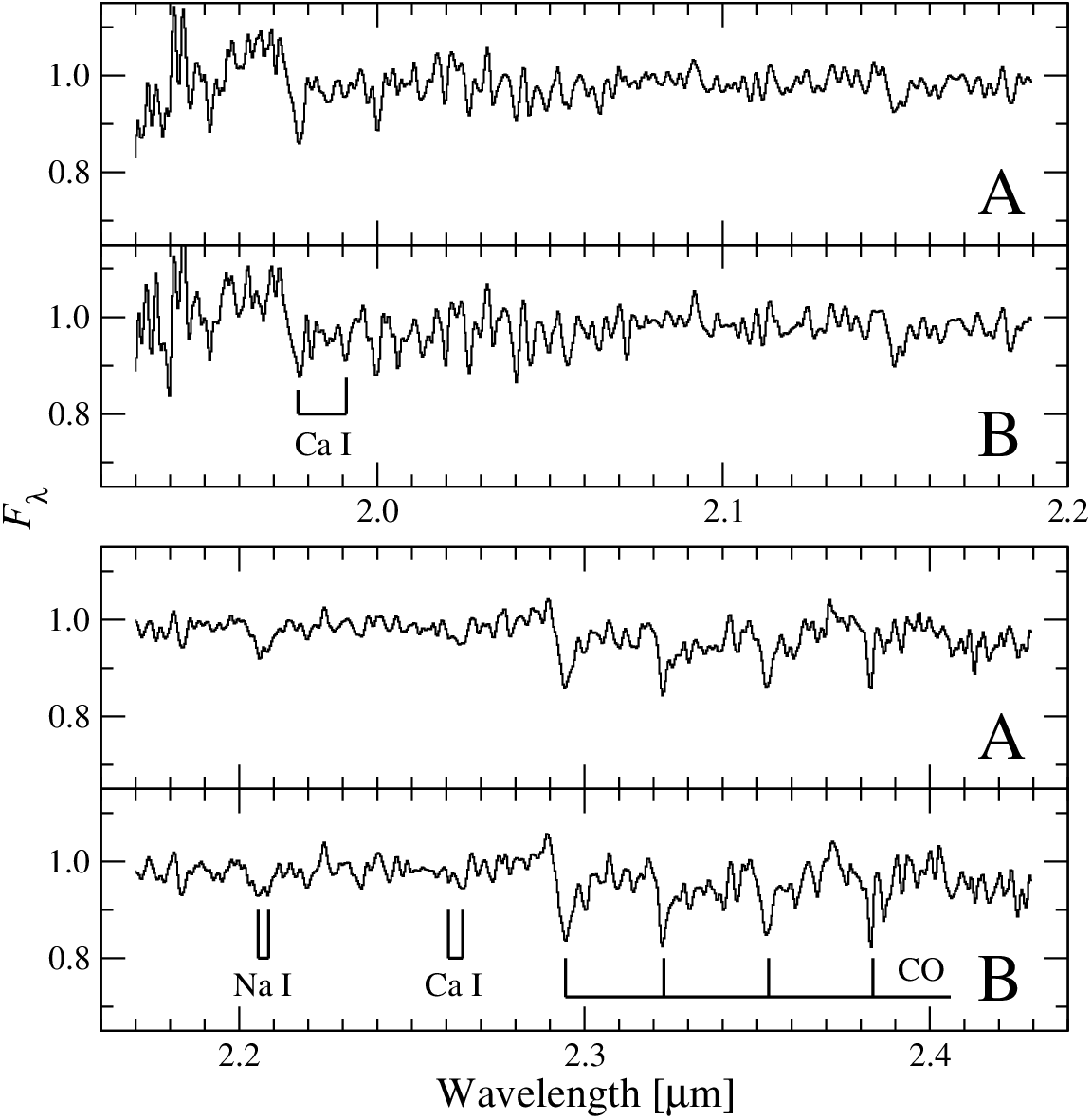}
\caption{\label{f:K} ISAAC $K$-band spectra of Oph\,1622 A and B, divided by the (pseudo-)continua. 
In addition to atomic features, the absorption train by molecular CO is identified.}
\end{figure}

\begin{figure}
\includegraphics[width=\hsize,clip=true]{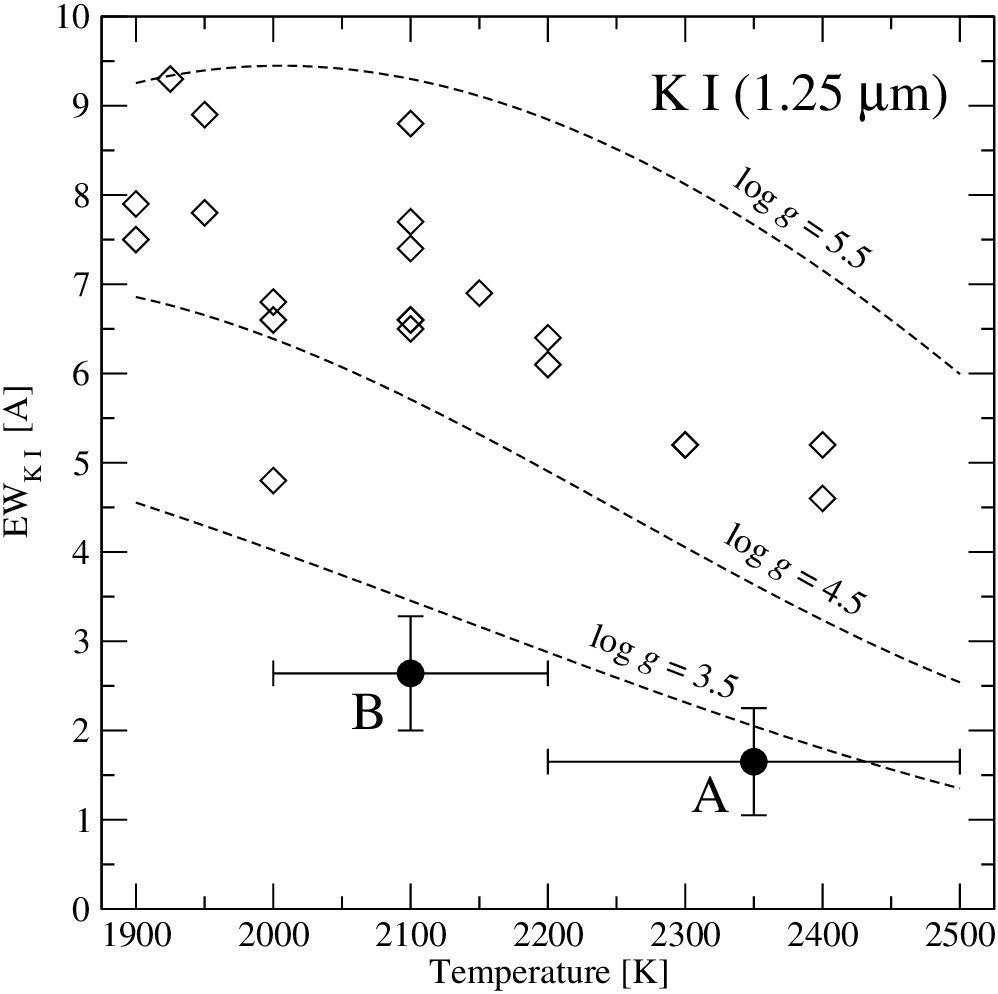}
\caption{\label{f:KI} (Pseudo-)equivalent widths of the gravity sensitive K\,I (1.25\,$\mu$m) line in Oph\,1622
A and B, compared to DUSTY models (dashed lines) and observations of field objects
\citep[diamonds, from][]{mcl03}. The estimated errors of the field objects are typically 
0.3\,\AA, and for the models 0.8\,\AA\ (coming from the uncertainty of the continuum level).
As is evident from this figure, the equivalent widths of Oph\,1622 A and B are far below the
expected values for field objects of the same spectral class, indicating low gravity and 
hence youth. Other K\,I lines in the spectra show the same trend.}
\end{figure}

\begin{deluxetable}{lcccc}
\tablewidth{0pt}
\tabletypesize{\scriptsize}
\tablecaption{\label{t:ew}Line equivalent widths}
\tablehead{
\colhead{Line} &
\colhead{EW$_{\mathrm{A}}$ [\AA]} &
\colhead{EW$_{\mathrm{B}}$ [\AA]} &
\colhead{$\lambda_0$ [$\mu$m]} &
\colhead{$\lambda_1$ [$\mu$m]}
}
\startdata
Na\,I (1.14\,$\mu$m)  & 5.0\,$\pm$\,1.6  &  6.1\,$\pm$\,1.5 & 1.137 & 1.140 \\
K\,I (1.17\,$\mu$m)  & 1.2\,$\pm$\,0.5  &  3.4\,$\pm$\,0.5 & 1.167 & 1.171  \\
K\,I (1.18\,$\mu$m)  & 2.5\,$\pm$\,0.5  &  4.0\,$\pm$\,0.4 & 1.176 & 1.179  \\
Fe\,I (1.19\,$\mu$m)  & 1.1\,$\pm$\,0.5  &  1.5\,$\pm$\,0.5 & 1.188 & 1.191 \\
K\,I (1.25\,$\mu$m)  & 1.6\,$\pm$\,0.6  &  2.6\,$\pm$\,0.6 & 1.251 & 1.253  \\
Na\,I (2.21\,$\mu$m)  & 4.3\,$\pm$\,0.8  &  4.4\,$\pm$\,0.9 & 2.201 & 2.213
\enddata
\tablecomments{The columns are the equivalent widths for component A and B, and the
integration interval used. The errors are estimated from the uncertainty in the 
continuum level.}
\end{deluxetable}

The merged SOFI spectra are shown in Fig.~\ref{f:sofi}, and the ISAAC $J$, $H$, and $K$
spectra in Figs.~\ref{f:J}--\ref{f:K}. The (pseudo-)equivalent width of several lines,
measured in the ISAAC spectra, are reported in Table~\ref{t:ew}. The separation of the binary was
measured from ISAAC spectra to be 1\farcs944$\pm$0\farcs010, where the estimate includes
uncertainties of the pixel scale (0.5\%) and orientation of the slit (1\degr). This agrees
well with previous measurements \citep{jay06c}.

Because of the great diversity of near-infrared spectral shapes found among young low-mass 
objects \citep{luc01,luc06}, and the lack of an extensive library of young late-type spectral standards,
it is difficult to constrain the spectral type from near-infrared spectra alone. The continuum
is suppressed by a multitude of molecular bands, many of which are gravity sensitive, and hence define
a `pseudo-continuum' that depends on both temperature and gravity. Finding a unique temperature and 
gravity by comparing to model atmospheres is also problematic; despite great advancements
of models in recent years, missing opacities limit their reliability on a detailed level
\citep[e.g.,][]{all00}. We compared the low-resolution SOFI spectra (Fig.~\ref{f:sofi}) to DUSTY
models by \citet{cha00} and more recent models by \citet[and priv.\ comm.]{bur06}, and found the 
data to be consistent with model temperatures between 
$T_{\mathrm{A}}=2200$--2500\,K for A, and $T_{\mathrm{B}}=2000$--2200\,K for B, similar to 
the temperatures estimated from optical spectra \citep{jay06c}. We were unable to constrain the 
gravity from the spectral shape, however, since the quality of the fit does not significantly change
between $\log g = 3.5$ and $\log g = 6.0$. 

Fortunately, there are a few atomic absorption features that appear to be
more gravity sensitive than the pseudo-continuum \citep{gor03}. In particular, the K\,I lines, seen in the 
ISAAC $J$-band spectra (Fig.~\ref{f:J}), are sensitive to both temperature and gravity. In Fig.~\ref{f:KI}
we present measurements of the equivalent width (EW) of the K\,I 1.25\,$\mu$m line in Oph\,1622 A and B,
with our temperature estimates. For comparison, we also plot the EW for field
dwarfs \citep[from][]{mcl03} and EWs measured in DUSTY synthetic spectra.
The major uncertainty in these measurements is the continuum, which is variable on small
scales due to molecular absorption bands. From the figure, it is clear that the K\,I EW of
Oph\,1622 A and B are significantly smaller than those of field dwarfs of similar atmosphere 
temperatures. According to the DUSTY models, a smaller EW is what we should expect for atmospheres 
of lower gravity. We therefore infer that the gravity is significantly lower for Oph\,1622 A and B than 
for field dwarfs, and that Oph\,1622 thus must be much younger than the field dwarfs. Repeating the
procedure for the other K\,I lines yields similar results. The EW of these and other lines in the 
spectra are reported in Table~\ref{t:ew}.

From evolutionary models by \citet{bar02}, one can derive the mass of an object once
either two of surface gravity, temperature and age are known. While the temperature is reasonably well
constrained, the surface gravity is not. A low-mass object of effective temperature 
$T$=2100--2500\,K is expected to have a surface gravity of $\log g \sim 3.6$ if the age is 
$\sim$1\,Myr, $\log g \sim 4.0$ if 5--10\,Myr, and $\log g \sim 5.3$ if in the field 
\citep{bar02}. Unfortunately, we find the DUSTY models to not be reliable enough at the 
level required to infer an accurate surface gravity. There seems to be a mismatch between 
the observed K\,I EWs for field objects from \citet{mcl03} and the expected EWs from models
(e.g., the K\,I 1.25\,$\mu$m line in Fig.~\ref{f:KI}, but also the other K\,I lines not shown). 
The degree to which this mismatch is reflected among lower gravity objects is not known, since we
lack a high-quality near-infrared spectroscopic library for young low-mass objects. In light of this
unknown systematic error, we do not attempt to derive an age from the measured gravity.
Instead, we assume that the age is comparable to other stars in the Ophiuchus star forming
region. Stars in the $\rho$\,Ophiuchi core region have ages 0.3--1\,Myr 
\citep{luh99}, while in the more extended Ophiuchus region ages have been estimated
to be 0.3--10\,Myr, with a median age of 2.1\,Myr \citep{wil05}. Combining our
temperature estimates with the age estimate from \citet{wil05}, we find that the 
\citet{bar02} models give the mass estimates  $M_{\mathrm{A}} = 13^{+8}_{-4}$\,M$_{\mathrm{J}}$ and 
$M_{\mathrm{B}} = 10^{+5}_{-4}$\,M$_{\mathrm{J}}$, where the main uncertainty is the assumed age
range (1--10\,Myr), and no systematic model uncertainty has been accounted for.

Our data are consistent with the near-infrared spectra of Oph\,1622 obtained independently 
by \citet{all06a} and \citet{clo06}. \citet{all06a} report temperatures of 2300/2200\,K and 
2800/2700\,K based on spectral energy distribution 
fitting and infrared spectral typing, respectively. However, the higher temperatures 
are not consistent with optical spectra \citep{jay06c}. Their much higher mass estimate of the binary 
($\sim$60/50\,M$_{\mathrm{J}}$) is a direct consequence of the high temperatures, also resulting in 
a high age estimate (40\,Myr) inconsistent with the region and the presence of circumstellar material
\citep{jay06c}. \citet{clo06} derive effective temperatures 2375/2175\,K and estimate masses to
$\sim$17.5/15.5\,M$_{\mathrm{J}}$, both consistent with our estimates. The reason 
that their mass estimates are slightly higher is that $\log g$ measured from gravity sensitive features are used 
directly, which is an independent method to constrain age and mass. As we have detailed 
above, however, it is also a difficult method, as a small error in $\log g$ results in a
large error in derived age (and mass).

\section{Conclusions}

\begin{enumerate}
\item The infrared spectra of Oph\,1622 A \& B are consistent with
  model atmospheres of temperatures $T_{\mathrm{A}}=(2350\pm150)$~K
  and $T_{\mathrm{B}}=(2100\pm100)$~K, confirming previous estimates from 
  optical spectra.
  
\item Model atmospheres together with the pseudo-continua in our 
  spectra do not significantly constrain the gravity of the atmospheres. Instead,
  we use gravity sensitive atomic K\,I absorption in the $J$-band and find 
  the equivalent widths to be consistent with significantly lower gravities than
  field dwarfs of the same spectral type, providing the best evidence so far
  that the Oph\,1622 binary is young.

\item Using our temperature estimates, the assumption that Oph\,1622 is of the same age
as other members of the Ophiuchus star forming region, and the \citet{bar02}
 evolutionary tracks, we derive component masses of $M_{\mathrm{A}} = 
 13^{+8}_{-4}$\,M$_{\mathrm{J}}$ and $M_{\mathrm{B}} = 10^{+5}_{-4}$\,M$_{\mathrm{J}}$,
 where the major uncertainty contribution comes from the age (assumed to be 1--10\,Myr).

\end{enumerate}

\acknowledgments
The observations reported here were collected under the ESO
program 277.C-5027.
We gratefully acknowledge the help from Marten H.\ van Kerkwijk
in getting his extraction routines to work, Aleks Scholz for general
support, Adam Burrows for computing and sending us model spectra,
and ESO for rapidly executing our DDT request. We made 
extensive use of NASA's Astrophysics Data System Bibliographic 
Services. This research was supported by an NSERC grant and
University of Toronto research funds to RJ.

{\it Facilities:} \facility{NTT (SOFI)}, \facility{VLT (ISAAC)}.


\begin{thebibliography}{31}
\expandafter\ifx\csname natexlab\endcsname\relax\def\natexlab#1{#1}\fi

\bibitem[{{Allard} {et~al.}(2000){Allard}, {Hauschildt}, \& {Schwenke}}]{all00}
{Allard}, F., {Hauschildt}, P.~H., \& {Schwenke}, D. 2000, \apj, 540, 1005

\bibitem[{{Allers}(2006)}]{all06a}
{Allers}, K.~N. 2006, Ph.D.~Thesis

\bibitem[{{Allers} {et~al.}(2006){Allers}, {Kessler-Silacci}, {Cieza}, \&
  {Jaffe}}]{all06b}
{Allers}, K.~N., et al. 2006, \apj, 644, 364

\bibitem[{{Baraffe} {et~al.}(2002){Baraffe}, {Chabrier}, {Allard}, \&
  {Hauschildt}}]{bar02}
{Baraffe}, I., et al. 2002, \aap, 382, 563

\bibitem[{{Baraffe} {et~al.}(2003){Baraffe}, {Chabrier}, {Barman}, {Allard}, \&
  {Hauschildt}}]{bar03}
{Baraffe}, I., et al. 2003, \aap, 402, 701

\bibitem[{{Barrado y Navascu{\'e}s} {et~al.}(2001){Barrado y Navascu{\'e}s},
  {Zapatero Osorio}, {B{\'e}jar}, {Rebolo}, {Mart{\'{\i}}n}, {Mundt}, \&
  {Bailer-Jones}}]{bar01}
{Barrado y Navascu{\'e}s}, D., et al. 2001, \aap, 377, L9

\bibitem[{{Bate} {et~al.}(2002){Bate}, {Bonnell}, \& {Bromm}}]{bat02b}
{Bate}, M.~R., {Bonnell}, I.~A., \& {Bromm}, V. 2002, \mnras, 332, L65

\bibitem[{{Bouy} {et~al.}(2003){Bouy}, {Brandner}, {Mart{\'{\i}}n}, {Delfosse},
  {Allard}, \& {Basri}}]{bou03}
{Bouy}, H., et al. 2003, \aj, 126, 1526

\bibitem[{{Burrows} {et~al.}(2006){Burrows}, {Sudarsky}, \& {Hubeny}}]{bur06}
{Burrows}, A., {Sudarsky}, D., \&  {Hubeny}, I. 2006, \apj, 640, 1063

\bibitem[{{Chabrier} {et~al.}(2000){Chabrier}, {Baraffe}, {Allard}, \&
  {Hauschildt}}]{cha00}
{Chabrier}, G., et al. 2000, \apj, 542, 464

\bibitem[{{Close} {et~al.}(2006){Close}, {Zuckerman}, {Song}, {Barman},
  {Marois}, {Rice}, {Siegler}, {Macintosh}, {Becklin}, {Campbell}, {Lyke},
  {Conrad}, \& {Le Mignant}}]{clo06}
{Close}, L.~M., et al. 2006, ApJ, submitted (astro-ph/0608574)

\bibitem[{{Cushing} {et~al.}(2003){Cushing}, {Rayner}, {Davis}, \&
  {Vacca}}]{cus03}
{Cushing}, M.~C., et al. 2003, \apj, 582, 1066

\bibitem[{{Gizis} {et~al.}(2003){Gizis}, {Reid}, {Knapp}, {Liebert},
  {Kirkpatrick}, {Koerner}, \& {Burgasser}}]{giz03}
{Gizis}, J.~E., et al. 2003, \aj, 125, 3302

\bibitem[{{Gorlova} {et~al.}(2003){Gorlova}, {Meyer}, {Rieke}, \&
  {Liebert}}]{gor03}
{Gorlova}, N.~I., et al. 2003, \apj, 593, 1074

\bibitem[{{Jayawardhana} {et~al.}(2003){Jayawardhana}, {Ardila}, {Stelzer}, \&
  {Haisch}}]{jay03}
{Jayawardhana}, R., et al. 2003, \aj, 126, 1515

\bibitem[{{Jayawardhana} \& {Ivanov}(2006{\natexlab{a}})}]{jay06c}
{Jayawardhana}, R., \& {Ivanov}, V.~D. 2006{\natexlab{a}}, Science, 313, 1279

\bibitem[{{Jayawardhana} \& {Ivanov}(2006{\natexlab{b}})}]{jay06b}
---. 2006{\natexlab{b}}, \apjl, 647, L167

\bibitem[{{Kraus} {et~al.}(2006){Kraus}, {White}, \& {Hillenbrand}}]{krau06}
{Kraus}, A.~L., {White}, R.~J., \& {Hillenbrand}, L.~A. 2006, ApJ, accepted (astro-ph/0602449)

\bibitem[{{Lucas} \& {Roche}(2000)}]{luc00}
{Lucas}, P.~W., \& {Roche}, P.~F. 2000, \mnras, 314, 858

\bibitem[{{Lucas} {et~al.}(2001){Lucas}, {Roche}, {Allard}, \&
  {Hauschildt}}]{luc01}
{Lucas}, P.~W., et al. 2001, \mnras, 326, 695

\bibitem[{{Lucas} {et~al.}(2006){Lucas}, {Weigths}, {Roche}, \&
  {Riddick}}]{luc06}
{Lucas}, P.~W., et al. 2006, MNRAS, accepted (astro-ph/0609086)

\bibitem[{{Luhman} {et~al.}(2005){Luhman}, {Adame}, {D'Alessio}, {Calvet},
  {Hartmann}, {Megeath}, \& {Fazio}}]{luh05}
{Luhman}, K.~L., et al. 2005, \apjl, 635, L93

\bibitem[{{Luhman} \& {Rieke}(1999)}]{luh99}
{Luhman}, K.~L., \& {Rieke}, G.~H. 1999, \apj, 525, 440

\bibitem[{{Mart{\'{\i}}n} {et~al.}(2001){Mart{\'{\i}}n}, {Zapatero Osorio},
  {Barrado y Navascu{\'e}s}, {B{\'e}jar}, \& {Rebolo}}]{mar01}
{Mart{\'{\i}}n}, E.~L., et al. 2001, \apjl, 558, L117

\bibitem[{{McLean} {et~al.}(2003){McLean}, {McGovern}, {Burgasser},
  {Kirkpatrick}, {Prato}, \& {Kim}}]{mcl03}
{McLean}, I.~S., et al. 2003, \apj, 596, 561

\bibitem[{{Moffat}(1969)}]{mof69}
{Moffat}, A.~F.~J. 1969, \aap, 3, 455

\bibitem[{{Pickles}(1998)}]{pic98}
{Pickles}, A.~J. 1998, \pasp, 110, 863

\bibitem[{{Rousselot} {et~al.}(2000){Rousselot}, {Lidman}, {Cuby}, {Moreels},
  \& {Monnet}}]{rou00}
{Rousselot}, P., et al. 2000, \aap, 354, 1134

\bibitem[{{Testi} {et~al.}(2002){Testi}, {Natta}, {Oliva}, {D'Antona},
  {Comeron}, {Baffa}, {Comoretto}, \& {Gennari}}]{tes02}
{Testi}, L., et al. 2002, \apjl, 571, L155

\bibitem[{{Wilking} {et~al.}(2005){Wilking}, {Meyer}, {Robinson}, \&
  {Greene}}]{wil05}
{Wilking}, B.~A., et al. 2005, \aj, 130, 1733

\bibitem[{{Zapatero Osorio} {et~al.}(2000){Zapatero Osorio}, {B{\'e}jar},
  {Mart{\'{\i}}n}, {Rebolo}, {Barrado y Navascu{\'e}s}, {Bailer-Jones}, \&
  {Mundt}}]{zap00}
{Zapatero Osorio}, M.~R., et al. 2000, Science, 290, 103

\end{thebibliography}

\end{document}